\documentclass[aps,twocolumn,groupedaddress,prb,floatfix,showpacs]{revtex4}
\usepackage{graphicx}% Include figure files
\usepackage{dcolumn}% Align table columns on decimal point
\usepackage{bm}% bold math
\usepackage{amsmath}% needed for subequations
\usepackage{verbatim}

\begin{document}

\title{Edge Saturation effects on the magnetism and band gaps in multilayer graphene ribbons and flakes}

\author{Bhagawan Sahu$^1$}
\email{brsahu@physics.utexas.edu}
\author{Hongki Min$^{2}$}
\author{Sanjay K. Banerjee$^1$}
\affiliation{
$^1$Microelectronics Research Center, The University of Texas at Austin, Austin, Texas 78758, USA\\
%$^2$Condensed Matter Theory Center, Department of Physics, University of Maryland, College Park, Maryland 20742, USA
$^2$Department of Physics, University of Maryland, College Park, Maryland 20742, USA
}

\date{\today}

\begin{abstract}

Using a density functional theory based electronic structure method and semi-local density approximation, we study the interplay of geometric confinement, magnetism and external electric fields on the electronic structure and the resulting band gaps of multilayer graphene ribbons whose edges are saturated with molecular hydrogen (H$_2$) or hydroxyl (OH) groups. We discuss the similarities and differences of computed features in comparison with the atomic hydrogen (or H-) saturated ribbons and flakes. For H$_2$ edge-saturation, we find \emph{shifted} labeling of three armchair ribbon classes and magnetic to non-magnetic transition in narrow zigzag ribbons whose critical width changes with the number of layers. Other computed characteristics, such as the existence of a critical gap and external electric field behavior, layer dependent electronic structure, stacking-dependent band gap induction and the length confinement effects remain qualitatively same with those of H-saturated ribbons.
\end{abstract}

\pacs{71.15.Mb, 81.05.ue, 73.22.Pr}

\maketitle

\section{Introduction}

Bulk graphene continue to attract considerable attention from the scientific community due to its novel electronic properties\cite{allan} and its potential applications in digital technologies\cite{sanjay} and in other niche areas\cite{rod}. The studies of finite size graphene has also increased in recent years. This is mainly due to advances in synthesis of sub-10 nm ribbons by lithographic techniques\cite{litho}, chemical methods\cite{chem} and unzipping of carbon nanotubes\cite{unzip}, identifying layer stacking order\cite{tony}, imaging of edge atoms one at a time\cite{kazu}, identification of edge types by atomic resolution technique\cite{novoselov}, the control of edge roughness with different edge saturations, probing edge magnetism\cite{mili}, and so on. Edge roughness is critical to the realization of graphene-based devices and recent advances in techniques focusing on nanoribbon growth process suggest edges with minimal or no roughness\cite{walter}. Recent density functional theory (DFT) studies of the energetics of edge adsorption and consequent change in the electronic spectrum in finite size ribbons, free of edge roughness, shows that the most energetically favorable edge saturation agents are oxygen followed by hydrogen\cite{vanin,ashwin}. These studies considered atomic (H) and molecular hydrogen (H$_2$), atomic oxygen (O), ammonia (NH$_3$), water(H$_2$O) and molecular nitrogen(N$_2$). DFT based calculations with local and semi-local functionals (local density approximation (LDA)\cite{ceperley} or gradient approximation(PW91)\cite{perdew} or generalized gradient approximation (PBE)\cite{burke} respectively) predict metallic nature for edge oxidized ribbons. NH$_3$ and N$_2$ binding to the edges was not possible with these semi-local functionals. However, edge saturations with atomic or molecular hydrogen and water seems to maintain graphene's gapped nature, opened by geometric confinements. We note here that edge saturation with hydrogen was used in recent experiments to probe the size dependent energy gaps in nanoribbons\cite{walter,lyding}. 

In this article, we report DFT based electronic structure calculations of multilayer graphene ribbons and flakes whose edges are saturated with molecular hydrogen (H$_2$) and the hydroxyl (OH) group. We will not discuss energetics of these edge saturations as it is addressed by various groups noted elsewhere in this section. Instead, we will focus on interplay of edge magnetism and the resulting electronic structure and the band gaps of ribbons and flakes with these saturations. This will help us to identify similarities and differences in their behaviors compared to those found for ribbons whose edges are saturated with atomic hydrogen\cite{sahu1,sahu2}. This work will provide comprehensive understanding of different edge saturation effects in nanoscale graphene fragments and will have implications in interpreting experiments.

The paper is organized as follows. In section II, we provide the details of the computational method and the parameters used for this study. The results of H$_2$ and OH-group saturations for the multilayer armchair ribbons will be presented in section III. We discuss the differences and similarities of our results with the published works on atomic hydrogen edge-saturated multilayer ribbons.  Section IV discusses the interplay of band gap and magnetism in zigzag ribbons with both type of saturations, and particularly the width-dependent magnetism in multilayer ribbons. Section V details our results for external electric field effects on band gaps. Length confinement effects on the band gaps will be discussed in Section VI. Finally, we conclude and summarize our results.          

\begin{figure}[ht!]
\includegraphics[width=1\linewidth]{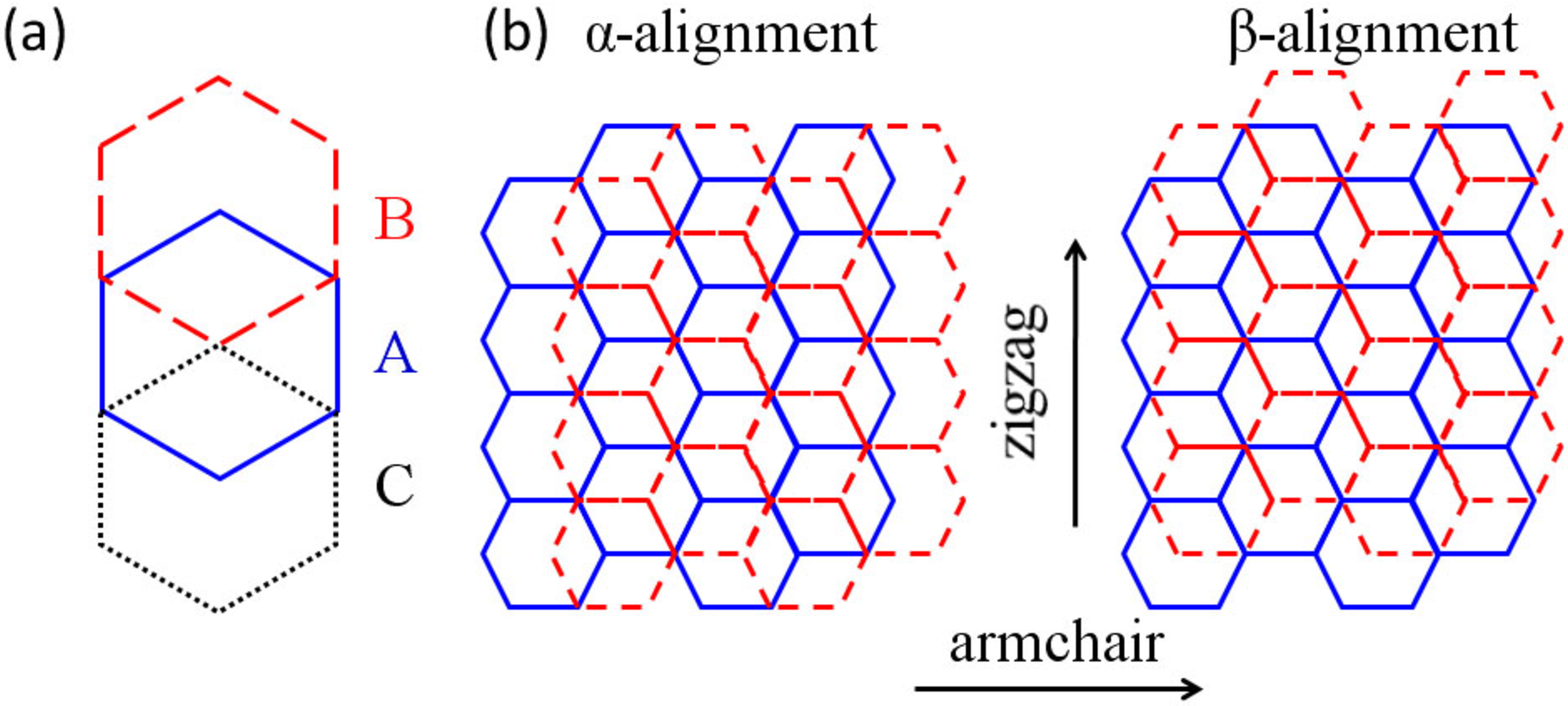}
\scalebox{0.38}{\includegraphics{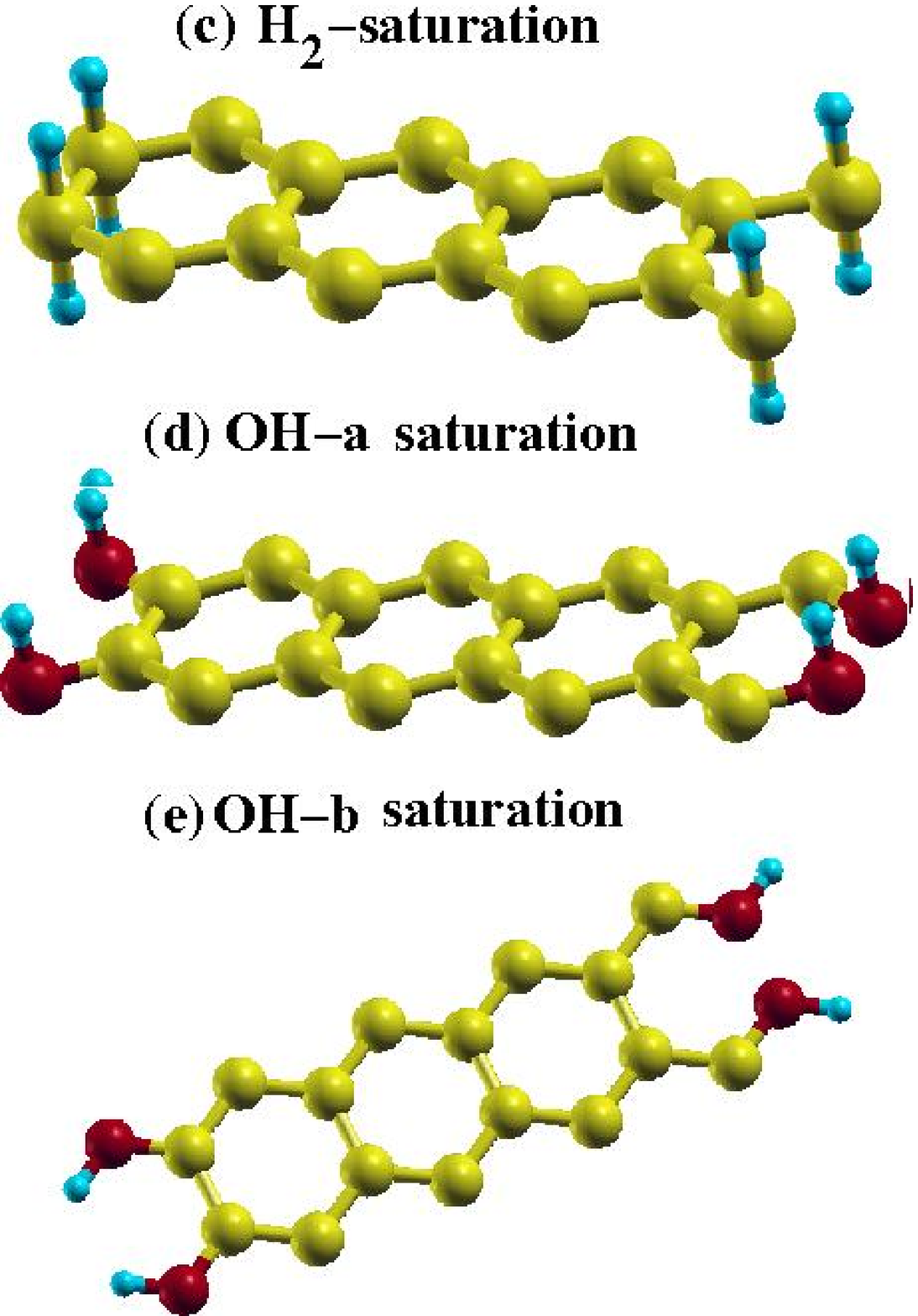}}
\caption{ (Color online) Schematic illustrations of (a) three types of stacking arrangements, labeled by A, B and C, (b) two types of edge alignments, $\alpha$-alignment and $\beta$-alignment in multilayer graphene nanoribbons. The two edge alignments are distinguished by different ways of shifting the top layer with respect to the other in finite size multilayer graphene stacks. The arrows indicate the direction of edges along which nanoribbons span infinitely. Monolayer graphene ribbon with (c) H$_2$ edge saturation and (d-e) two different possibilities of OH-group edge saturation. In the OH edge-saturated ribbons, hydrogen can lie in perpendicular to the plane of graphene (d) or be parallel to it (e), denoted as OH-{\it a} and OH-{\it b}, respectively. The color scheme chosen for the atoms are: carbon (yellow), oxygen(red) and hydrogen (blue). We also consider these edge saturations for bilayer and multilayer ribbons. 
}
\label{fig:Fig1}
\end{figure}

\section{Computational Method and the Edge Saturated Structures}

We use an electronic structure method \cite{paolo} implemented within DFT, with ultrasoft pseudopotentials \cite{ultrasoft} for core-valence interactions and the plane-wave basis, to obtain graphene ribbon and flake  band structures with different edge saturations. Our previous studies\cite{sahu1,sahu2} of multilayer ribbons and flakes with atomic hydrogen saturation suggests that the interlayer distance and the appearance of edge magnetism is sensitive to the particular local or semi-local approximation used. Therefore, for the sake of consistency and meaningful comparisons with the published works on atomic hydrogen saturation, the gradient approximation (GGA)\cite{perdew} is used in this study to capture the edge magnetism with the fixed interlayer distance of 0.335 nm. The van der Waal's interaction, which anchors the graphene layers, is not included in our calculations as it is shown to have weak influence on overall band structure at a given distance\cite{kronik,timo}. The unsaturated carbon $\sigma$-orbitals were passivated with H$_2$ and OH-group atoms (Fig. 1). We relaxed the oxygen, hydrogen and carbon atomic positions in all graphene fragments considered in this study. No such relaxations were necessary in atomic hydrogen-saturated ribbons or flakes as the initial C-H distance (taken from CH$_4$ molecule and set at 0.1084 nm) and the resulting electronic structure was found to be insensitive to atomic relaxations.  We chose d(C-H)=0.1084 nm, d(C-O)=0.1262 nm (taken from CO molecule) and d(O-H)$\sim$0.1 nm (taken from water molecule) as initial distances which changed to new values by atomic relaxations.   

The ribbons were placed in a supercell with vacuum regions adjacent to the width and the stacking direction to make it an isolated system in DFT-based calculations. For monolayer and bilayer graphene, we used 1 nm and 1.5 nm vacuum region, respectively, whereas for more than two layers, larger vacuum sizes were considered. Different vacuum size in the calculations guaranteed that the periodic images of the supercell do not interact with each other. For graphene flake calculations, vacuum regions were considered in all three crystallographic directions. We used 68 {\bf k}-points in the irreducible part of the Brillouin Zone (BZ) for ribbon and 8 {\bf k}-points for flake calculations. For both ribbons and flakes, kinetic energy cut-off of 475 eV was used. The convergence of the calculations were tested with respect to a denser {\bf k}-point mesh, larger energy cut-offs, as well as larger vacuum sizes. For atomic relaxations, the calculations were assumed to be converged when the maximum Hellman-Feynman force components were less than the chosen threshold 0.01 V/nm. The convergence was tested with stricter force cut-off values.

To establish the ground state magnetic order for the zigzag ribbons, we tested both narrow and wide ribbons with non-magnetic, collinear (ferromagnetic and antiferromagnetic) and non-collinear order between the layers, while we set ferromagnetic coupling along each edge and antiferromagnetic coupling between the two edges within the same layer as a starting magnetic configuration, as predicted theoretically for monolayer graphene ribbons\cite{fujita}. We find that interlayer antiferromagnetic order has lower energy than ferromagnetic, non-magnetic, or non-collinear magnetic order in the ribbons with both H$_2$ and OH-group edge saturations. We also find that the same ground state is reached with other forms of semi-local exchange-correlation potentials such as PBE \cite{burke}, PBESol \cite{perdew1}, RPBE \cite{zhang}. Therefore, for calculating band structures and other related quantities in this article, we consider interlayer antiferromagnetic order as a magnetic ground state of multilayer graphene ribbons and flakes with both molecular and OH edge saturations.

\section{Armchair graphene sheets}

In this section, we address the interplay of band-gap and edge saturations with H$_2$ and OH-group in multilayer armchair ribbons and compare the results with atomic hydrogen edge saturation. First we discuss the results for H$_2$ saturation followed by OH-group saturation.  
   
\subsection{Molecular Hydrogen and OH-group saturation}

We consider multilayer armchair graphene ribbons with widths as large as 5 nm for this study whose edges are saturated with H$_2$ (Fig. 1(c)). The most energetically favored layer stacking sequences in multilayer graphene stacks was chosen: Bernal (or AB) stacked ribbons for bilayer graphene and flakes and both AB- as well as ABC-periodic stacks for multilayer ribbons. Our study suggests three classes of ribbons with H$_2$ saturation. However, there is one important difference compared to the ribbon classes with atomic hydrogen edge saturations: the number of armchair rows {\it N} defining a ribbon width and the {\it triad} of ribbon widths forming a set containing three classes is shifted. For example, for ribbons with H edge saturation, {\it N}=3$p$+2 results in a lowest gap within a given {\it class}, while for H$_2$ saturation it is the ribbon with {\it N}=3$p$+1 that has the lowest gap where $p$ = 1, 2, 3,..... is an integer. We call these ribbons {\it metallic}, though for small widths, the ribbons are actually semiconducting. This shift in width values associated with H$_2$ edge saturation can be understood by invoking the configurational changes to the initial hydrogen position (set perpendicular to the graphene plane) after the system reaches relaxed ground state suggested by our calculations: the C-H bond and angle change from the initial value of 0.1084 nm and 180$^o$ to 0.1114 nm and 100$^o$. This bending results in interaction of $\pi$-orbitals of graphene on the outermost edge and hydrogen orbitals resulting in rearrangement of the hybridized states. It modifies $sp^2$ hybridization to $sp^3$-like hybridization destroying bare $\pi$ orbitals at the outermost edges and therefore, shift the width values. Figure 2(a) shows the variation of gap values with widths of monolayer graphene ribbons. The behavior is similar to that found for ribbons with atomic hydrogen edge saturation: decreasing gaps with increasing width values and the class distinctions becoming weaker for wide ribbons. Compared to the gaps for atomic hydrogen edge-saturated ribbons, the gaps in ribbons with H$_2$ are consistently smaller. Bilayer ribbons with $\alpha$-edge alignments (Fig. 1(b)) also show three classes, albeit with smaller gaps compared to the monolayer gaps, a behavior similar to the atomic hydrogen saturated monolayer and bilayer ribbons (Fig. 2(b)). Our studies also predict three classes for bilayer armchair ribbons with $\beta$-edge alignment as well as multilayer armchair ribbon stacks (Figures not shown).    

\begin{figure}[ht!]
\scalebox{0.47}{\includegraphics[angle=0]{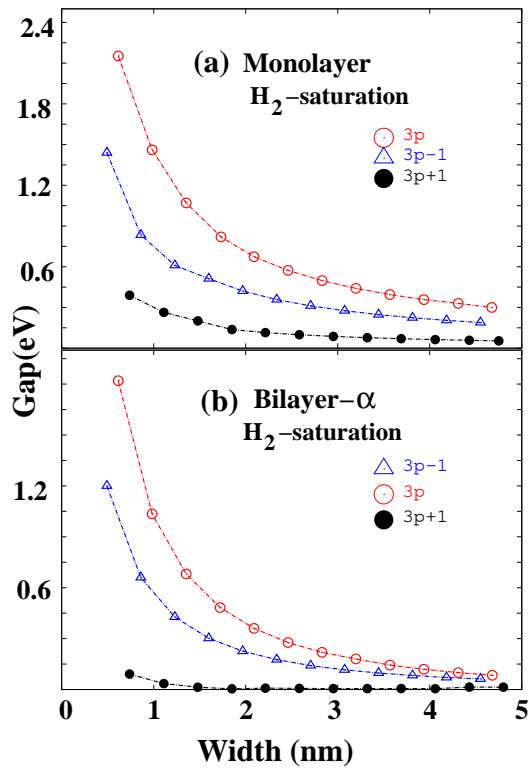}}
\caption{(Color online) Width variation of gap values in armchair ribbons with H$_2$ edge saturation in (a) monolayer and (b) bilayer graphene with $\alpha$-edge alignments. The three classes of ribbons with {\it shifted} labeling of armchair rows defined by {\it N}=3$p$, 3$p$-1 and 3$p$+1, where $p$ = 1, 2, 3,..... is an integer, is shown. The bilayer ribbons with $\beta$-alignment as well as multilayer ribbons show similar behavior.}    
\label{fig:Fig2}
\end{figure}

Figure 3 shows band structures of trilayer armchair ribbons stacked in ABA and ABC fashion. For metallic ribbons, our study suggests one linear and one quadratic band near the Fermi level (placed at zero) for ABA stacked ribbons (Fig 3(a)) whereas we find one cubic band for ABC stacked ribbon (Fig 3(b)). These predictions are consistent with the low energy states and energy dispersions near the Fermi level in ribbons with atomic hydrogen edge saturations\cite{sahu1} as well as in bulk multilayer graphene\cite{hongki}.   

Now we discuss the armchair ribbons with OH-group saturation. There are two possible initial configurations suggested for this group in the literature\cite{cho}. They differ in a way the hydrogen atom is positioned with respect to the graphene plane: perpendicular to the plane (Fig. 1(d)) or parallel to it ( Fig. 1(e)). For later discussions, we refer to these configurations, respectively as, OH-{\it a} and OH-{\it b}. 

\begin{figure}[ht!]
\scalebox{0.390}{\includegraphics{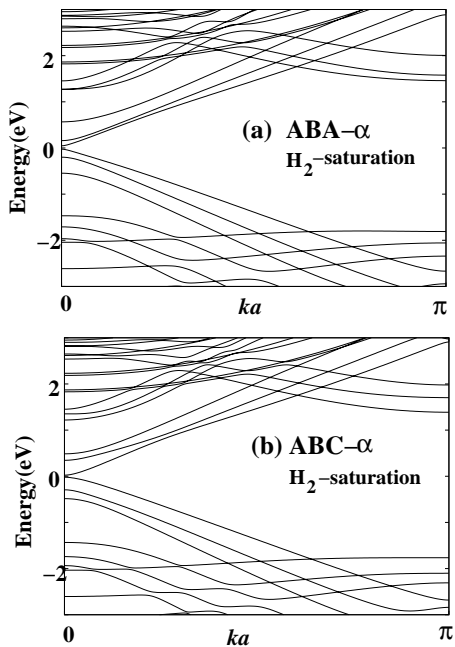}}
\caption{ The band structures of trilayer graphene nanoribbon with edges saturated with H$_2$ and $\alpha$-edge alignments between the layers for (a) ABA layer stacking and (b) ABC layer stacking sequence. The ribbons with width values corresponding to {\it N}=7 are chosen. The low energy dispersion of the states near the Fermi level is consistent with atomic hydrogen edge saturation as well as the bulk multilayer graphene.
}
\label{fig:Fig3}
\end{figure}

\begin{figure}[ht!]
\scalebox{0.390}{\includegraphics{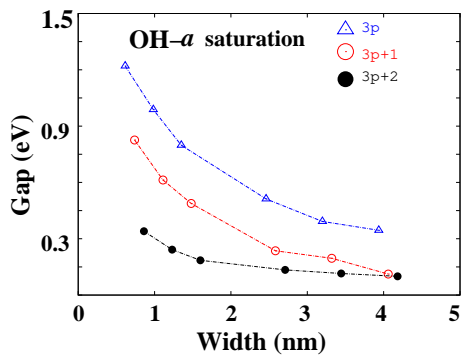}}
\caption{ (Color online) The band gap variation with the width of monolayer armchair ribbons whose edges are saturated with OH-{\it a} group. Three classes of ribbons are seen with the labeling as those with atomic hydrogen saturation but the size of the induced gap for the ribbons with labels 3$p$ and 3$p$+1, in a given {\it class}, are reversed where $p$ = 1, 2, 3, ... is an integer. No such reversal is found for monolayer ribbons with OH-{\it b} group edge saturations, while bilayer and multilayer ribbons with both OH-{\it a} and OH-{\it b} group-saturations exhibit the gap reversal trend. The definition of OH-{\it a} and OH-{\it b}, are given in the text.
}
\label{fig:Fig4}
\end{figure}

Both the OH-group configurations, for wide and narrow monolayer and bilayer graphene ribbons, maintain the planar and non-planar arrangement of hydrogen atoms in the final equilibrium ground state reached with the atomic relaxations. The C-O and O-H bond distances change to 0.14 nm (0.133 nm) and 0.103 nm (0.097 nm) for OH-{\it a} (OH-{\it b}) configurations respectively and the angles change to 121$^o$ (112$^o$) for OH-{\it a} (OH-{\it b}) configurations respectively. However, our study suggests that the non-planar configuration (OH-{\it a}) is energetically favorable (the energies differ by at least 0.2 eV/atom for both wide and narrow ribbons) compared to the planar OH-{\it b} configuration. We find no significant differences in their electronic spectrum.      

Both the OH-configurations show three ribbon classes in monolayer graphene. For ribbons with OH-{\it a} edge saturation, {\it N}=3$p$ class has the largest energy gap compared to {\it N}=3$p$+1 and {\it N}=3$p$+2 classes, while for ribbons with OH-{\it b} saturations, it is the {\it N}=3$p$+1 class which shows the largest energy gap, similar to those with atomic hydrogen saturations. Here $p$ = 1, 2, 3, ..... is an integer. Figure 4 show the gap variation of monolayer ribbon with the edges saturated with OH-{\it a} group which clearly show the trend. Moreover,the gaps in this case is larger than those of ribbons saturated with H or H$_2$. The bilayer and multilayer ribbons with both OH-{\it a} and OH-{\it b} group-saturation also exhibit three classes of ribbons with the gap reversal trend predicted for monolayer ribbons with OH-{\it a} edge saturation (Figures not shown). Our calculations suggest that the band structures of metallic ribbons with OH-group edge saturation follow the dispersion trends seen in ribbons with either atomic hydrogen or H$_2$ edge-saturation or bulk multilayer graphene namely the existence of low energy states near the Fermi level, decomposed into states whose wave-vector dependence is dictated by the layer stacking sequences.   

\section{Zigzag graphene sheets}

\begin{figure}[ht!]
\scalebox{0.45}{\includegraphics[angle=0]{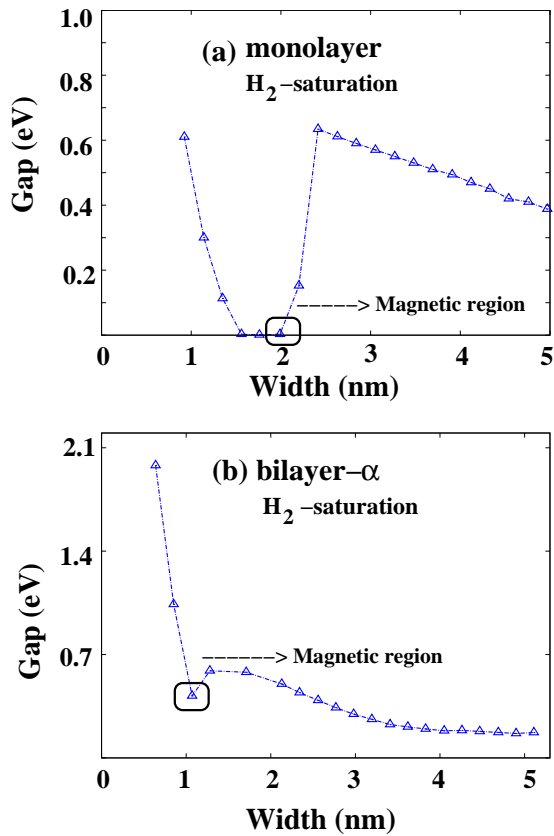}}
\caption{ (Color online) The width dependent magnetism in (a) monolayer and (b) bilayer zigzag ribbons with H$_2$ edge saturation. The magnetic to non-magnetic transition is found at a critical width of 2 nm for monolayer and 1 nm for bilayer ribbons. The non-magnetic to magnetic transition point is marked with a square and in both cases magnetic regions are clearly indicated.}
\label{fig:Fig5}
\end{figure}

We discuss edge magnetism and resulting band gaps in multilayer zigzag graphene ribbons saturated with H$_2$ and the OH-group. Our studies suggests width dependent magnetism in ribbons with H$_2$ edge saturation. Figure 5(a) and (b), respectively, shows the variation of gap values with the width of the monolayer and bilayer zigzag ribbons. Monolayer ribbons with widths as large as 2 nm or more are found to be magnetic whereas our studies predict magnetism for very narrow width bilayer zigzag ribbons (with widths $\sim$1 nm or less) and narrowest multilayer ribbons are magnetic. To understand the nature of this magnetic to non-magnetic transition, we plot band structures of monolayer zigzag ribbons with few representative widths: above (W = 2.5 nm) and below (W = 1.1 nm) and at the critical width of 2 nm. Figure 6(a) shows the non-magnetic ribbon band structure with the width W = 2.5 nm. A degenerate flat-band occur at the Fermi level hinting at instability towards magnetism. When we consider magnetic order in our calculation, a magnetic ground state is realized which induces a finite gap in the energy spectrum at the Fermi level (Fig. 6(b)). For the widths below the critical width (W = 1.1 nm), no such degenerate flat bands close to Fermi level are seen (Figs. 6(e)), which make the ribbons with these widths non-magnetic.  A finite gap is already visible in the non-magnetic spectrum near the Fermi level which is due to increasing geometrical confinement with decreasing widths. Considering magnetic order has no effect on the non-magnetic nature of the ribbon as well as on its gap value (Fig.6(f)). Bilayer ribbons show similar flat bands near the Fermi level for widths above 1 nm which drives the system towards magnetism (Figures not shown). This hints at the interplay of quantum confinement, magnetism and edge saturation in graphene nanoribbons. We note here that ribbons with atomic hydrogen edge saturation and OH-group saturation do not show such width dependent magnetism in the graphene stack with any number of layers.

\begin{figure}[ht!]
\scalebox{0.35}{\includegraphics[angle=0]{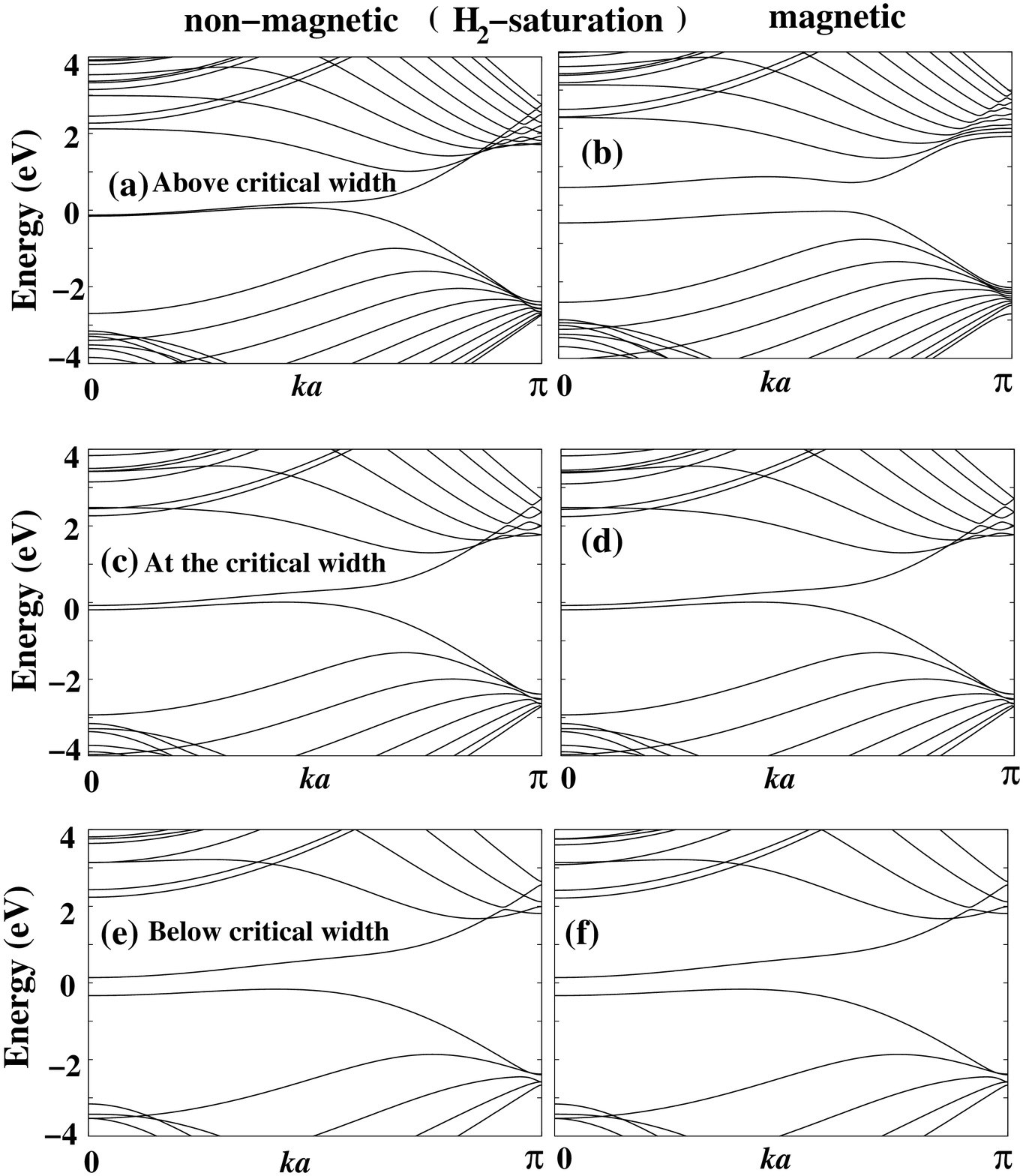}}
\caption{ The band structures of monolayer zigzag ribbons with H$_2$ edge saturations at the width of 2.5 nm with (a) non-magnetic and (b) magnetic calculations, at the critical width of 2.0 nm with (c) non-magnetic and (d) magnetic calculations and at the width of 1.1 nm with (e) non-magnetic and (d) magnetic calculations which are, respectively, above, at and below the critical width of 2 nm predicted for magnetic to non-magnetic transitions. The nature of this transition, with ribbon widths, is related to the existence of flat bands at the Fermi level. Bilayer ribbons show a similar transition, albeit at lower widths.}
\label{fig:Fig6}
\end{figure}

\begin{figure}[ht!]
\scalebox{0.35}{\includegraphics[angle=0]{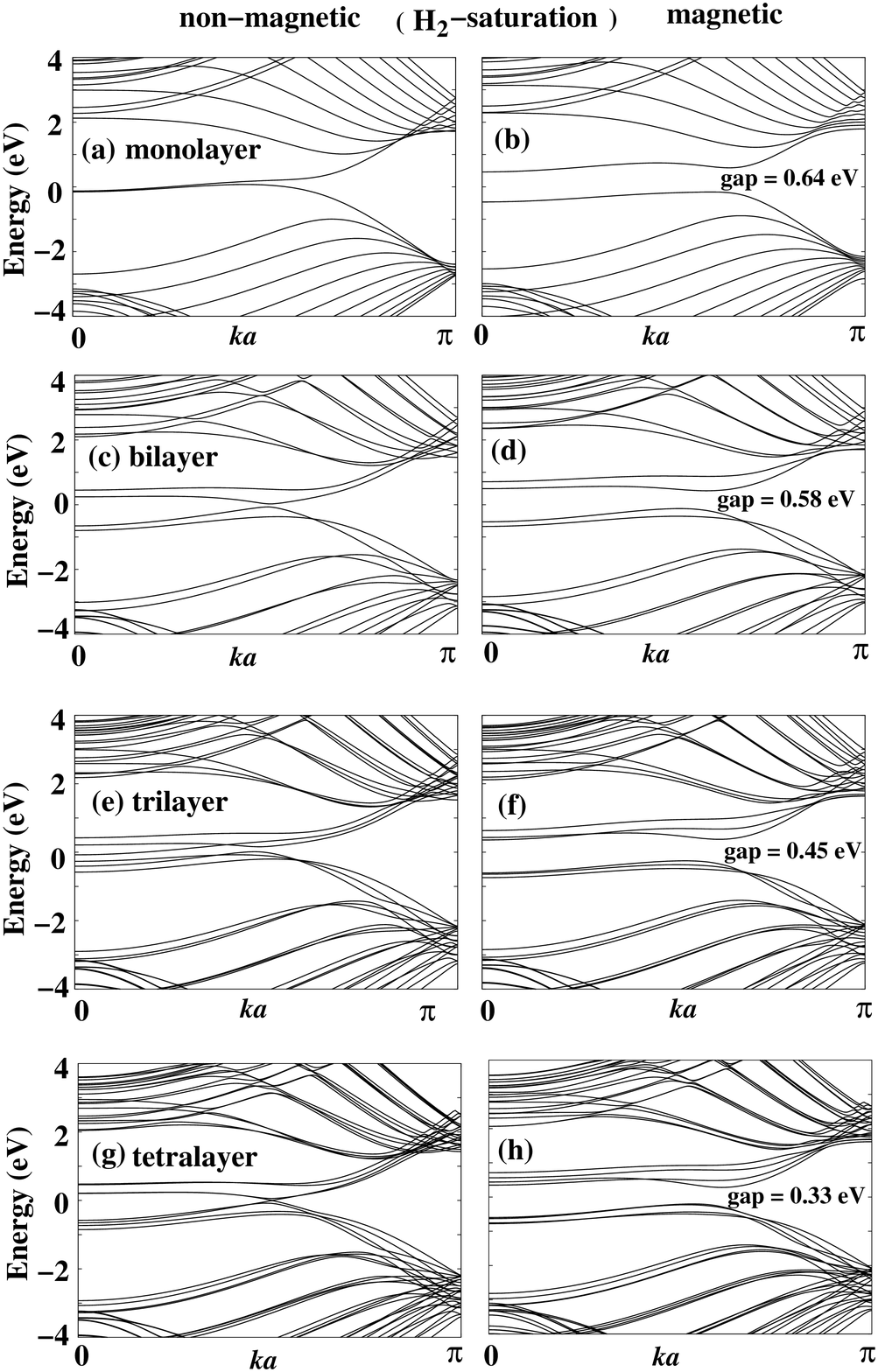}}
\caption{ The band structures of multilayer zigzag ribbons with H$_2$ edge saturation exhibiting layer number dependent features at or near the Fermi level in the non-magnetic ground state for the odd layer numbered ribbons (a and e) and even layer numbered ribbons (c and g). In odd layer numbered ribbons such as mono and trilayer ribbons, flat bands occur at the Fermi level whereas for bilayer and tetralayer ribbons, it occurs away from the Fermi level. The size of the induced gap, due to edge magnetism, depend on these features.
}
\label{fig:Fig7}
\end{figure}

We predict layer number dependent electronic structures in multilayer ribbons with H$_2$ and OH-group edge saturations. The results are similar as found for multilayer ribbons with atomic hydrogen saturations\cite{sahu2}. Since both the H$_2$ and OH-group saturations provide similar results, we discuss our results for ribbons with H$_2$ edge-saturation only. Figure 7 shows the band structures for multilayer zigzag ribbons with upto four layers. Both non-magnetic and magnetic ribbons were considered. In non-magnetic ribbons with odd number of layers (monolayer and trilayer ribbons), the flat bands populate the Fermi level in most of the BZ (Figs. 7(a) and (e)), but in ribbons with even number of layers (bilayer and tetralayer ribbons) (Figs. 7(c) and (g)), the flat bands lie close to the Fermi level with only a very small portion occupying the BZ. Therefore, ribbons with both odd and even number of layers are expected to be unstable towards the magnetic ground state. Considering layer and edge magnetism in our calculations results in magnetic ground state  which is lower in energy by about 0.1 eV from the non-magnetic ground state and induces a band gap (Figs 7(b), (d), (f) and (h)). Since odd layered ribbons have more flat bands at the Fermi level compared to even layered ribbons, the induced gap due to edge magnetism is expected to be larger in the former. But we do not find such increase in gap sizes in ribbons with odd number of layers compared to that with even number of layers. The ribbons with OH group edge saturation show similar odd-even effects (Figures not shown).  

\begin{figure}[ht!]
\scalebox{0.45}{\includegraphics[angle=0]{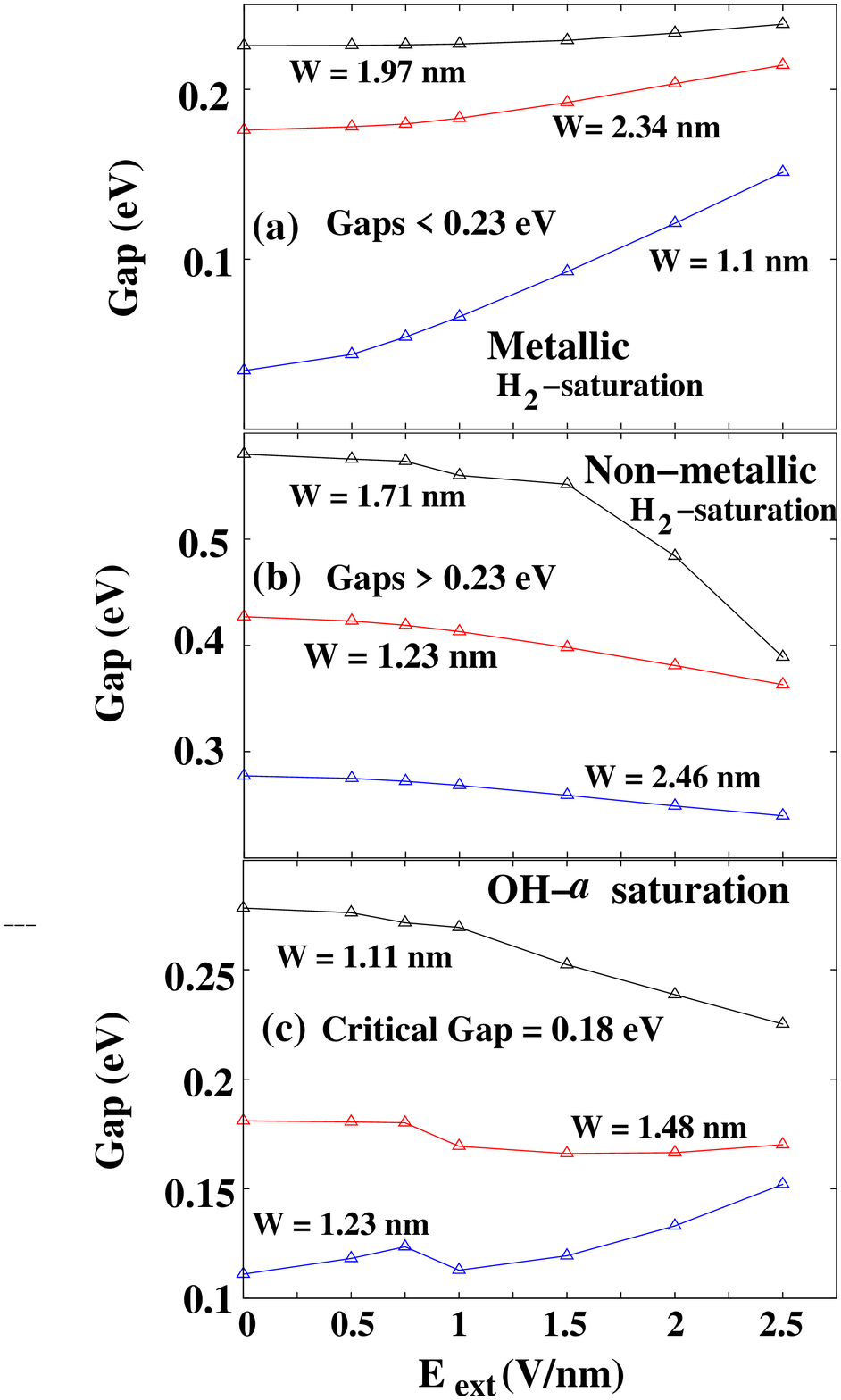}}
\caption{ (Color online) The external electric field effects on the gap values of (a) metallic armchair and (b) non-metallic armchair and zigzag bilayer ribbons with H$_2$ edge-saturation and (c) for ribbons with OH-{\it a} edge saturation. The metallic ribbons initially with the small gap show increase in the gap values with increase of external field strengths whereas non-metallic ribbons initially with large gap values show decreasing trend in all cases. A critical gap value of 0.23 eV  (0.18 eV)is predicted for ribbons with H$_2$ (OH-{\it a}) edge saturation above and below which electric field effects have opposite sign.
}
\label{fig:Fig8}
\end{figure}

\begin{figure}[ht!]
\scalebox{0.45}{\includegraphics[angle=0]{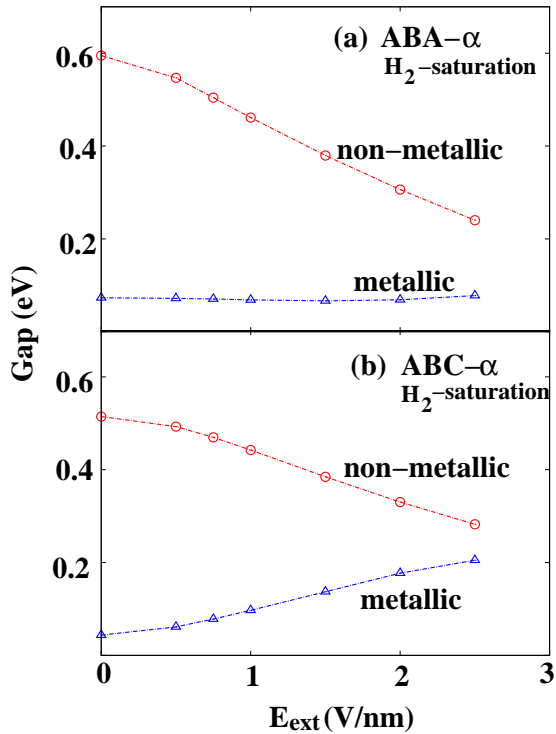}}
\caption{ (Color online) The external electric field effects on the gap values of trilayer metallic (width W = 0.74 nm) and non-metallic (W = 0.86 nm) armchair ribbon with H$_2$ edge-saturation. In ABC-stacked trilayer metallic ribbons, the gap values are enhanced (a) and it remain constant or closed in ABA-stacked ribbons (b) which is consistent with those predicted for bulk multilayer graphene and ribbons with atomic hydrogen edge saturations. The non-metallic ribbon gaps decrease with increasing electric field strengths.
}
\label{fig:Fig9}
\end{figure}

\begin{figure}[ht!]
\scalebox{0.35}{\includegraphics[angle=0]{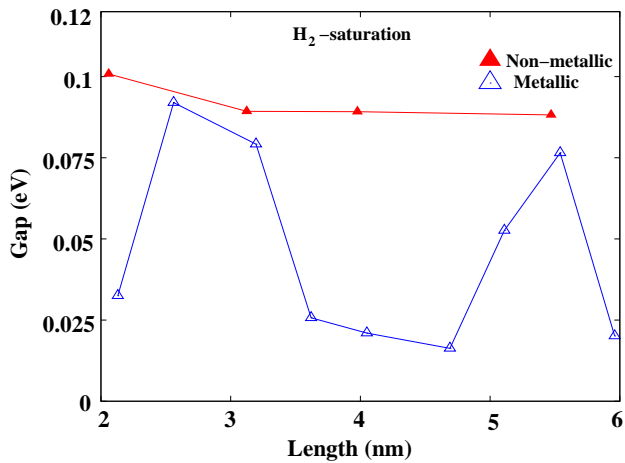}}
\caption{ (Color online) The length confinement effects on the gap values of bilayer armchair metallic and non-metallic ribbons  whose edges are saturated with H$_2$. For metallic ribbons, the length has significant effect on the gap values. 
}
\label{fig:Fig10}
\end{figure}

\section{External Electric Field Effects}

This section deals with effects of perpendicular electric field on the band gap of bilayer and multilayer graphene ribbons with H$_2$  and OH-group edge saturation. The saw-tooth type of electric potential was used for electric field studies and electric field values up to 2.5 V/nm is applied to both wide and narrow gap ribbons. The choice of the maximum electric field values is guided by our recent studies of bulk trilayer\cite{fan} and finite size multilayer ribbons\cite{sahu1}. Few representative widths, W= 1.1 nm, 1.23 nm, 1.97 nm, 2.34 nm and 2.5 nm with initial gap values of 0.035 eV, 0.43 eV, 0.226 eV, 0.176 eV and 0.277 eV were chosen in order to locate a {\it critical} gap below (above) which gaps increase (decrease) with electric field, which may exist as in multilayer ribbons\cite{sahu1,sahu2} with atomic hydrogen saturation. Figure 8 shows gap values versus the electric field strengths for bilayer armchair ribbons whose edges are saturated with H$_2$. Our studies suggests a critical gap of 0.23 eV, which is quite close to 0.2 eV predicted for bilayer and multilayer ribbons with atomic hydrogen saturations. This indicates relative insensitiveness of the value of the {\it critical} gap with respect to different edge saturations. For ribbons with gap values below the critical gap, the electric field has the effect of increasing it (Fig. 8(a)) and the opposite behavior is seen for ribbons with gap above the critical gap (Fig. 8(b)). The critical gap value is lowered to 0.18 eV (Figure 8(c)) in ribbons with OH-{\it a} saturation.

Figure 9 shows the variation of gap values in ABA and ABC layer stacked trilayer ribbons with $\alpha$-edge alignment. Both 
metallic and semiconductor ribbons were considered. In metallic ABC-periodic ribbons, a gap is induced upon application of
external electric fields which is consistent with the gap opening in ABC-periodic bulk trilayer graphene\cite{fan} as well as metallic trilayer ribbons with atomic hydrogen edge saturation\cite{sahu1}. The electric field effect on the gap values of ribbons with OH group edge saturation exhibits similar behavior (Figures not shown).  

\section{Length effects}

In this section we discuss the length confinement effect on the gap values of bilayer ribbons with H$_2$ edge-saturation. Bilayer ribbons were considered as a representative case study so that we can compare our results of bilayer ribbons with atomic hydrogen edge saturation\cite{sahu2}. Both wide (or small gap or metallic) and narrow (or large gap or semiconducting) width ribbons were considered with length values as large as 6 nm. Figure 10 shows gap values versus the lengths for both metallic and non-metallic armchair ribbons. Metallic ribbons show strong variations in the gap values compared to the non-metallic ribbons with respect to change in the length values. The non-monotonic behavior of the gap values with change in metallic ribbon length is quite interesting. For monolayer\cite{nayak} and bilayer ribbons\cite{sahu2} with atomic hydrogen edge saturations, such behavior was predicted and it was suggested that the different behavior of metallic and non-metallic ribbons originate from the electronic states which are localized along the width (or zigzag) direction in the later but completely delocalized throughout the flake in the former. We believe that this will also be true for ribbons with edge saturations other than atomic hydrogen except for the fact that the degree of localization or delocalization is now determined by the hybridization of carbon $\pi$-orbitals with the edge saturations. Confining the atomic hydrogen-saturated ribbons along the length direction (armchair or zigzag) increases the gap values due to magnetism compared to the non-magnetic ribbons\cite{sahu1}. Our studies suggests that in ribbons with H$_2$ edge-saturation, no such clear increase of gap values is seen probably due to relative positions of hydrogen atoms with respect to the carbon $\pi$-orbitals.              

\section {Summary and Conclusions}
We use a density functional based electronic structure method to study the effect of different edge saturations on the magnetism and band gaps of multilayer ribbons and flakes. Two energetically favorable saturations namely H$_2$ and OH-group are considered. Our study suggests both distinct and similar electronic properties for both the armchair and zigzag ribbons and flakes, compared to the atomic hydrogen edge saturation. For H$_2$ saturation, we predict {\it different} labeling of the three armchair ribbon classes and the width dependent magnetism in zigzag ribbons whose critical width for magnetic to non-magnetic transition changes with number of layers, while for OH-group saturation, the ribbons qualitatively follow the characteristics of H-saturated ribbons.  We identify a {\it critical} gap above and below which external electric field has opposite behavior. Length confinement studies suggest significant gap variations in metallic ribbons. Our studies have implications for interpreting future experiments and provide understanding of interplay of magnetism and band gaps in finite size graphene with different edge saturations.         

\acknowledgments
The authors acknowledge financial support from Nanoelectronics Research Initiative supported Southwest Academy of Nanoelectronics (NRI-SWAN) center. We thank the Texas advanced computing center (TACC) for computational support on {\it Ranger} (TG-DMR080016N) and {\it Lonestar}.

\end{document}